\documentclass[aps,prd,preprint,groupaddress,amssymb,amsmath,nofootinbib]{revtex4}

\usepackage{color}
\usepackage{graphicx}
\usepackage[usenames,dvipsnames]{xcolor}

\begin{document}

\preprint{}

\title{
Unification of Gauge Couplings in the Standard Model with Extra Vector-like Families }

\author{Radovan Derm\' \i\v sek}

\affiliation{Physics Department, Indiana University, Bloomington, IN 47405, USA}



\date{February 14, 2013}

\begin{abstract}

We discuss gauge coupling unification in  models with additional  1 to 4 complete vector-like families, and derive simple rules for masses of vector-like fermions required for exact gauge coupling unification. These mass rules and the classification scheme are generalized to an arbitrary extension of the standard model. 
We focus on scenarios with 3 or more vector-like families in which the values of  gauge couplings at the electroweak scale  are highly insensitive to the grand unification scale,  the unified gauge coupling, and the masses of vector-like fermions. Their observed values can be mostly understood from  infrared fixed point behavior. With respect to sensitivity to fundamental parameters, the model with 3 extra vector-like families stands out. It requires  vector-like fermions with masses of order 1 TeV -- 100 TeV, and thus at least part of the spectrum may be within the reach of  the LHC. The constraints on proton lifetime can be easily satisfied in these models since the best motivated grand unification scale is  at $\sim$$10^{16}$ GeV. The Higgs quartic coupling remains positive all the way to the grand unification scale, and thus the electroweak minimum of the Higgs potential is stable.  

\end{abstract}


\maketitle






\section{Introduction}

Models for new physics at the TeV scale are typically motivated by the hierarchy problem. They strive to explain the hierarchy between the electroweak (EW)  scale and the Planck scale, or at least remove  the incredible fine tuning required in the standard model (SM) for having such a hierarchy. However, the SM is stubbornly surviving the first tests at the LHC and there are no traces of new physics yet. In addition, the mass of the recently discovered Higgs-like particle suggests that the SM can be a consistent theory all the way to the Planck scale. This gives more weight to speculations that there is no mechanism, no new physics,  that stabilizes the hierarchy, or that the EW scale is selected based on anthropic reasoning.

However, even when we ignore the hierarchy problem, the SM is still not very satisfactory. The three gauge couplings, all couplings of the Higgs boson to fermions, the Higgs mass, and the Higgs quartic coupling are free parameters. This motivates us to explore extensions of the standard model  in which at least some of these parameters could be understood.

We have recently showed that extending the standard model by three complete vector-like families (SM+3VFs) with masses of order  1 TeV - 100 TeV allows for the unification of gauge couplings~\cite{Dermisek:2012as}. Predictions for gauge couplings at the EW scale are highly insensitive to fundamental parameters: the grand unification scale,  the unified gauge coupling, and the masses of vector-like fermions. Their observed values can be mostly understood from  infrared fixed point behavior.

In this paper we discuss gauge coupling unification in detail in  models with additional  1 to 4 complete vector-like families (VFs), and derive simple rules for masses of vector-like fermions required for exact gauge coupling unification. We then focus on scenarios with 3 or more vector-like families that lead to  insensitive unification of gauge couplings. Requiring the smallest splitting between masses of vector-like fermions we show that the best motivated grand unified theory (GUT) scale is  at $\sim$$10^{16}$ GeV. We provide examples of the spectrum as a function of the GUT scale, which can be as large as the Planck scale. We discuss constraints from proton decay and show that predictions from the best motivated region are close to current limits.  However, due to insensitivity of predicted EW scale values of gauge couplings to GUT scale parameters, no sharp predictions can be made without knowing the spectrum of vector-like fermions.

The focus on complete families follows from the fact that quantum numbers of quarks and leptons in the SM  nicely fill  representations of a GUT symmetry, $\bf 10$ and $\bf \bar 5$ of $SU(5)$ or $\bf 16$ of $SO(10)$. This provides a support for the idea of grand unification and the unification of gauge couplings~\cite{PDG_GUTs}. Additional complete families
represent some of the simplest  extensions of the SM  that can be embedded into simple GUTs.\footnote{This does not mean that the masses of vector-like fermions needed for gauge coupling unification necessarily result from a simple unified boundary condition. By simple GUTs we mean that there is no additional mechanism required to keep particles  in incomplete GUT multiplets significantly below the GUT scale, or to split their masses over  many orders of magnitudes that would,  to large extent, ameliorate the motivation for GUTs.} Consequently, there are many studies exploring various features of vector-like families (mostly in supersymmetric models), see for example Refs.~\cite{Babu:1996zv, BasteroGil:1999dx, Kolda:1996ea, Barr:2012ma, Martin:2009bg}. 

In addition, vector-like fermions, not necessarily coming in complete GUT multiplets, are often introduced on purely phenomenological grounds to explain various discrepancies between observations and SM predictions. Examples include discrepancies in precision EW Z-pole observables~\cite{Choudhury:2001hs, Dermisek:2011xu, Dermisek:2012qx, Batell:2012ca}, and  the muon g-2 anomaly~\cite{Kannike:2011ng}. However, with arbitrary new particles  there are many possibilities for gauge coupling unification.\footnote{For examples of recent studies investigating the effects of extra particles on gauge coupling unification in models without supersymmetry, see Refs.~\cite{Kopp:2009xt, Giudice:2012zp}.} Therefore, we generalize the mass rules and the method to classify scenarios consistent with gauge coupling unification  to an arbitrary extension of the standard model. 

The method to classify scenarios consistent with gauge coupling unification in terms of physical masses of extra particles starts with finding the mass scales  that represent ``average" masses  of all particles charged under given gauge symmetry  required for gauge coupling unification (they are defined precisely in the next section and are referred to as crossing scales). These crossing scales are easy to obtain and they  immediately give us information about the required spectrum. First of all, if they do not exist between the EW scale and the GUT scale, the gauge coupling unification in a given model is not possible, no matter what the splitting between masses of extra particles is. Second, the splitting between  crossing scales represents the minimum necessary splitting in the spectrum required. Third, from the mass formulas that define crossing scales in terms of masses of extra particles one can immediately see the basic features of the spectrum required, and the spectrum can be calculated. In addition, these formulas also indicate the freedom one has in imposing further relations between masses of extra particles. This might be useful when searching  for  models that relate masses of particles at a given scale. The mass rules given in terms of particle masses can be evolved to an arbitrary scale, {\it e.g.} the GUT scale, which would provide the boundary conditions that need to be satisfied. However, the renormalization group (RG) evolution of the mass rules  depends on additional assumptions one has to make about the origin of the masses and the scale at which these masses are generated.\footnote{The study of gauge coupling unification is, to large extent, unaffected by  these assumptions; only the physical masses of particles  matter in the leading order. If the masses originate from Yukawa couplings to extra scalars that get  vacuum expectation values at an intermediate scale, these may contribute to the RG evolution of gauge couplings at 2-loop level. However unless the extra couplings are large these effects would be negligible.}

This paper is organized as follows. In Sec.~\ref{sec:RGE} we discuss RG evolution of gauge couplings  in models with extra VFs. We start with the discussion of IR fixed point predictions for gauge couplings, then add  threshold corrections from a universal mass of vector-like fermions, and, finally, we add effects from splitting masses of vector-like fermions. We discuss sensitivity of predicted values of gage couplings to fundamental parameters. Finally, we derive simple mass rules that have to be satisfied in order to get exact gauge coupling unification. We generalize the method to classify all solutions consistent with gauge coupling unification to an arbitrary extension of the SM. In Sec.~\ref{sec:discussion} we discuss constraints from proton decay, the stability of the EW minimum of the Higgs potential, and discuss a possible origin of masses of vector-like fermions. We give few concluding remarks in Sec.~\ref{sec:conclusions}.

\section{Renormalization group evolution of gauge couplings}
\label{sec:RGE}

The one-loop renormalization group
equations (RGEs) for three gauge couplings, $\alpha_i = g_i^2 / 4\pi$, are given by:
\begin{equation}
\frac{d \alpha_i}{dt} \; = \beta(\alpha_i) \; = \; \frac{\alpha_i^2}{2 \pi} \,  b_i,
\label{eq:RGE}
\end{equation}
where $t = \ln Q/Q_0$ with $Q$  representing the energy scale at which gauge couplings are evaluated. The beta function coefficients, $b_i$, in the SM with $n_f$ families are given by
\begin{equation}
 b_i \; = \; \left( \frac{1}{10}+\frac{4}{3}n_f , \;   - \frac{43}{6} +\frac{4}{3}n_f, \; -11 + \frac{4}{3}n_f\right).
 \label{eq:bi}
\end{equation}
For $n_f = 3$ we get the usual SM result, $b_i = (41/10, \; -19/6, \; -7)$. With  extra $N$ pairs of complete VFs we have $n_f = 3 + 2\times N $
(a vector-like partner contributes in the same way).  For example, in SM+3VFs we find $b_i = (121/10, \; 29/6, \; +1)$ which indicates that all three gauge couplings are asymptotically divergent (this result obviously holds for  3 or more pairs of VFs). 

\begin{figure}[t]
\includegraphics[width=3.in]{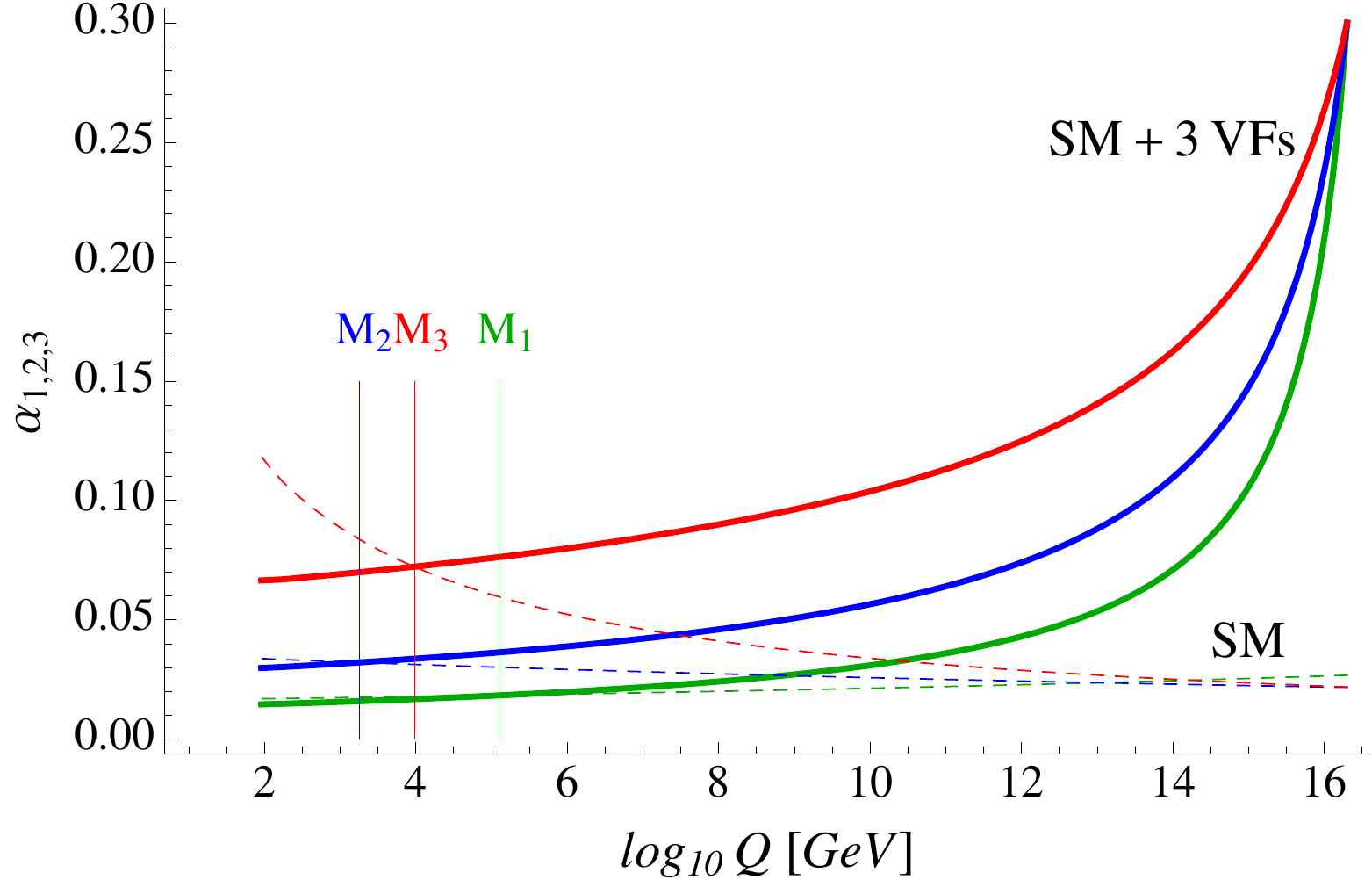}
\includegraphics[width=3.in]{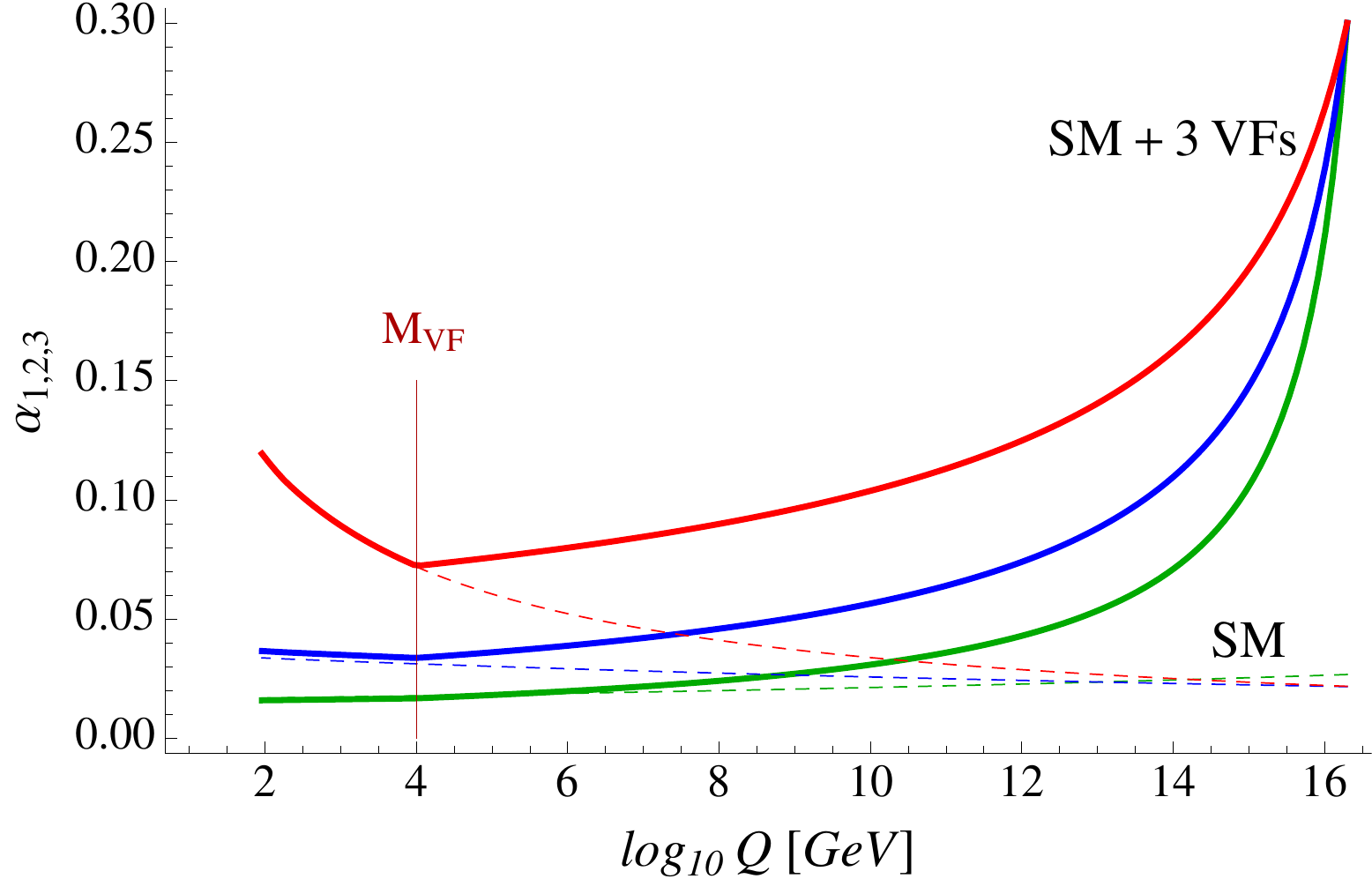}
\caption{RG evolution of gauge couplings: $\alpha_3$ (top solid line), $\alpha_2$ (middle solid line), and $\alpha_1$ (bottom solid line) in the SM extended by three vector-like families for $\alpha_G = 0.3$ at $M_G = 2 \times 10^{16}$ GeV.  Dashed lines in the same order  show  the running of gauge couplings in the SM.   Masses of  3 VFs are neglected in the left plot, and fixed to 10 TeV (indicated by $M_{VF}$) in the right plot.
  The crossing points in the evolution of gauge couplings in the SM+3VFs and the SM indicated in the left plot define the common threshold scales, $M_{1,2,3}$, for masses of particles charged under given symmetry required for exact gauge coupling unification. }
\label{fig:thresholds}
\end{figure}

The evolution of gauge couplings in the SM  and an example of the evolution  in the SM+3VFs case are shown in Fig.~\ref{fig:thresholds}. The numerical analysis closely follows that of Ref.~\cite{Dermisek:2012as}. For the SM evolution we use the Z-scale 
central values of $\alpha_{EM}^{-1} (M_Z) = 127.916$, $\sin^2 \theta_W = 0.2313$, and $\alpha_{3} (M_Z) = 0.1184$, together with the top quark mass $m_t = 173.2$ GeV, which can be found in Ref.~\cite{Nakamura:2010zzi}. The $\alpha_{EM}$ and $\sin^2 \theta_W$ are related to $\alpha_{1,2} (M_Z)$ through 
 \begin{eqnarray}
\sin^2 \theta_W &=&  \frac{\alpha'}{ \alpha_2 +  \alpha'} , \quad {\rm and} \quad \alpha_{EM}  \; = \; \alpha_2 \, \sin^2 \theta_W,
\end{eqnarray}
where, assuming the $SU(5)$ normalization of the hypercharge, $\alpha' \equiv  (3/5)  \alpha_1$. We set the Higgs boson mass to $m_h = 126$ GeV~\cite{:2012gk, :2012gu}. 
The example of the RG evolution of gauge couplings in the SM+3VFs starts with unified gauge coupling $\alpha_G = 0.3$ at $M_G = 2 \times 10^{16}$ GeV. The crossing points in the evolutions of gauge couplings in these two cases, which will be important for the discussion of threshold corrections, are indicated in the left plot by $M_{1,2,3}$.
In all numerical results  we use full two loop RGEs~\cite{Machacek:1983tz}, we integrate out  all particles with masses above $M_Z$  at their mass scale, and include one-loop matching corrections for $m_t$ and $m_h$~\cite{Hambye:1996wb}.  We assume that Yukawa couplings of vector-like fermions are negligible, and we also neglect Yukawa couplings of all fermions in the SM except the top quark.

The results of the numerical analysis  we present can be understood from approximate analytic formulas.
The one-loop RGEs can be  solved, and we can express   gauge couplings at the EW scale  in terms of the GUT scale,  and  values of  gauge couplings at $M_G$:
 \begin{equation}
 \alpha_i^{-1} (M_Z) \; = \;  \frac{b_i}{2 \pi}  \ln \frac{M_G}{M_Z} \; + \;  \alpha_i^{-1} (M_G) .
 \label{eq:sol_1loop}
  \end{equation}
 Assuming gauge coupling unification, $\alpha_i (M_G) = \alpha_G$, and neglecting threshold corrections both at the EW scale and the GUT scale, we can  express one gauge coupling in terms of the other two. For example:
   \begin{equation}
 \alpha_3 (M_Z) \; = \;  \frac{b_1 - b_2}{(b_1 - b_3) s_W^2  + 3/5 (b_3 - b_2) c_W^2}  \, \alpha_{EM}  (M_Z),
 \label{eq:prediction_1loop}
  \end{equation}
where $s_W^2 \equiv \sin^2 \theta_W (M_Z)$, and $c_W^2 \equiv \cos^2 \theta_W (M_Z)$. For the measured values of $\alpha_{EM}$ and $s_W^2$  the SU(5) embedding of the SM predicts 
 $\alpha_3 (M_Z) \simeq 0.07$ which is about 40\% below the experimental value.

 Adding complete chiral or vector-like families at the EW scale does not change at all the one-loop prediction given in Eq.~(\ref{eq:prediction_1loop}) since complete families contribute equally to all three beta function coefficients, see Eq.~(\ref{eq:bi}). Furthermore, the scale of unification (more precisely the scales where any two couplings meet) does not change at one-loop, only the value of the unified gauge coupling increases. With increasing the number of extra families, at some point, the couplings become non-perturbative before they meet, and eventually reach the Landau pole. Further increase of the number of families  lowers the energy scale at which the Landau pole occurs.

However, the SM extended with a sufficient number of complete vector-like families  so that all couplings  are asymptotically divergent offers a new possibility. Vector-like families  introduce an additional scale to the problem associated with  masses of vector-like fermions, $M_{VF} $, and they contribute to the RG evolution of gauge couplings only above this energy scale. This allows us to consider models with a large (but still perturbative) unified gauge coupling at a high scale, higher than the scale at which the Landau pole would occur if the VFs were at the EW scale. Consequently, in the RG evolution to lower energies, gauge couplings run to the (trivial) infrared (IR) fixed  point.  Thus, at lower energies, the values of gauge couplings are determined only by the particle content of the theory and how far from the GUT scale we measure them. Since the exact value of $\alpha_G$ becomes irrelevant, instead of one prediction of the conventional unification, Eq. (\ref{eq:prediction_1loop}), we have two predictions for ratios of gauge couplings. At the $M_{VF} $ scale, the vector-like fermions are integrated out, and below this scale  gauge couplings run according to the usual RG equations of the standard model. In a way, the two parameters of the conventional unification, $M_G$ and $\alpha_G$, are replaced by $M_G$ and $M_{VF} $. The discrepancies of IR fixed point predictions from observed values can be explained by threshold effects of extra vector-like fermions.

\subsection{IR fixed point predictions for gauge couplings}

The IR fixed point predictions were discussed in detail in Ref.~\cite{Dermisek:2012as}. In models with asymptotically divergent couplings, these can be easily obtained if the 1-loop RGEs are good approximations. 
Assuming a large enough unification scale and large (but still perturbative) unified gauge coupling, the first term in Eq.~(\ref{eq:sol_1loop}) dominates,
  and the ratios of gauge couplings are given by ratios of beta function coefficients, 
   \begin{equation}
 \frac{\alpha_i (M_Z)}{  \alpha_j (M_Z)} \; \simeq \;  \frac{b_j} {b_i} .
\end{equation}
This can be translated  into the prediction for $\sin^2 \theta_W$:   
 \begin{equation}
\sin^2 \theta_W \equiv  \frac{\alpha'}{ \alpha_2 +  \alpha'} = \frac{b_2}{b_2+b'},
\label{eq:s2w}
\end{equation}
where 
$b' \equiv  (5/3) b_1$.
Numerically, we find $ \sin^2 \theta_W  =  0.193$ in the case of SM+3VFs, which is identical to the value obtained assuming 9 chiral families~\cite{Maiani:1977cg, Cabibbo:1982hy}.
Similarly, in SM+4VFs we find $ \sin^2 \theta_W  =  0.234$.

In the case of SM+3VFs, the one-loop RGE for  $\alpha_3$ given in Eq.~(\ref{eq:RGE}) is not a good approximation because of the accidentally small $b_3$ coefficient. The two-loop contribution to the beta function is well approximated by the term proportional to  $\alpha_3^3$, 
\begin{equation}
\frac{d \alpha_3}{dt} \; = \beta(\alpha_3) \; \simeq \; \frac{\alpha_3^2}{2 \pi} \,  b_3 \; +\; \frac{\alpha_3^3}{8 \pi^2} \,  B_3,
\label{eq:RGE2l}
\end{equation}
where $B_3 = -102 + (76/3) n_f = 126$ for SM+3VFs~\cite{Machacek:1983tz}. 
Thus, the two loop contribution is larger than the one-loop contribution for $\alpha_3 \gtrsim 0.1$.\footnote{This is a consequence of a very small 1-loop beta function coefficient and it is not an indication of non-perturbativity. The coupling is still perturbative, the dominant 3-loop contribution to $\beta(\alpha_3)$,  proportional to $\alpha_3^4$,  represents a $\sim 5 \%$ correction to 1 + 2-loop beta function for $\alpha_3 \simeq 0.1$~\cite{Mihaila:2012fm}.}

The RGE for $\alpha_3$ can be solved by  adding the 1-loop contribution  as an expansion in $\epsilon = 4 \pi b_3/B_3$ to the solution obtained from the 2-loop contribution only~\cite{Grunberg:1987sk}, \cite{Dermisek:2012as}. Alternatively, we can solve the full RGE given in Eq.~(\ref{eq:RGE2l}) and find:
 \begin{equation}
 \alpha_3^{-1} (M_Z)  - \frac{1}{\epsilon} \ln \left(1 + \frac{\epsilon}{\alpha_3 (M_Z) }\right)\; = \;  \frac{b_3}{2 \pi}  \ln \frac{M_G}{M_Z} + \alpha_G^{-1}  - \frac{1}{\epsilon} \ln \left(1 + \frac{\epsilon}{\alpha_G }\right) .
  \label{eq:sol_2loop}
  \end{equation}
Neglecting $\alpha_G^{-1}$, we obtain the second  prediction:
   \begin{equation}
\frac{ \alpha_3 (M_Z)}{1 - \frac{\alpha_3 (M_Z)}{\epsilon} \ln ( 1 + \frac{\epsilon}{\alpha_3 (M_Z) })} \; = \;  \frac{b_2+b'}{b_3}   \, \alpha_{EM}  (M_Z).
 \label{eq:prediction_1+2loop}
  \end{equation}
Numerically, for  $ \alpha_{EM}  (M_Z) = 1/127.916$, it predicts  $ \alpha_3 (M_Z)  \simeq 0.072$ in the case of SM+3VFs.

 The beta function coefficients for $\alpha_3$ in the SM+4VFs scenario are $b_3 = 11/3$ and $B_3 = 530/3$. The 1-loop term in the RG equation (\ref{eq:RGE2l}) dominates for $\alpha_3 < 0.26$ in this case.

The proximity of predictions from the IR fixed point, Eqs.~(\ref{eq:s2w}) and (\ref{eq:prediction_1+2loop}), to observed values is certainly intriguing.
Although they are not a perfect match to  measured values,  the discrepancies can be easily accommodated by taking into account  threshold corrections from 
vector-like fermions  that should be integrated out at the $M_{VF}$ scale.

 \subsection{Mass scale of vector-like fermions and sensitivity to fundamental parameters}
 
The existence of a scale associated with masses of vector-like fermions  is strongly suggested by the overlay of the RG evolution of gauge couplings in the SM and those in the SM+3VFs assuming unified gauge coupling at a high scale given in Fig.~\ref{fig:thresholds}. All three gauge couplings in these two scenarios cross at comparable scales suggesting a common threshold at which particles from VFs are integrated out. Indeed, for the example given in Fig.~\ref{fig:thresholds}, fixing all the masses  of 3 VFs to 10 TeV, shown in Fig.~\ref{fig:thresholds} (right), the EW scale values of gauge couplings are predicted within 8\% from measured values. In the next subsection, we will show that the measured values  of gauge couplings can be precisely reproduced by splitting the masses of vector-like fermions. First, however, we would like to discuss general features of this result assuming the common mass of VFs.

The fairly good agreement of predicted values of gauge couplings from 3 VFs at $\sim 10$ TeV with observed values does not rely on the specific choice of the GUT scale and the value of the unified gauge coupling. The EW scale values of  gauge couplings are  highly insensitive to these parameters, which can be understood from IR fixed point behavior.

\begin{figure}[t]
\includegraphics[width=2.9in]{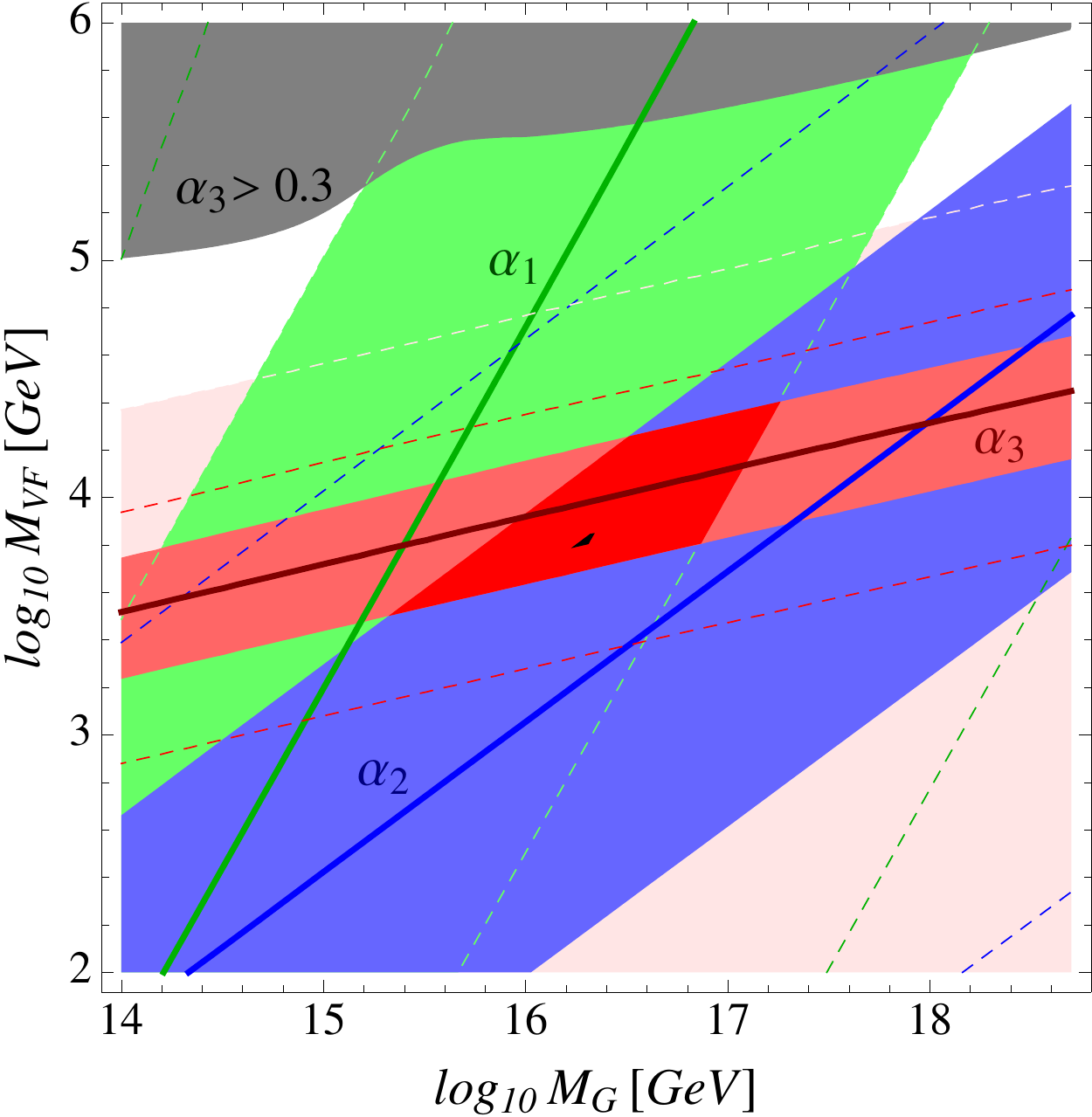} \hspace{0.5cm}
\includegraphics[width=2.9in]{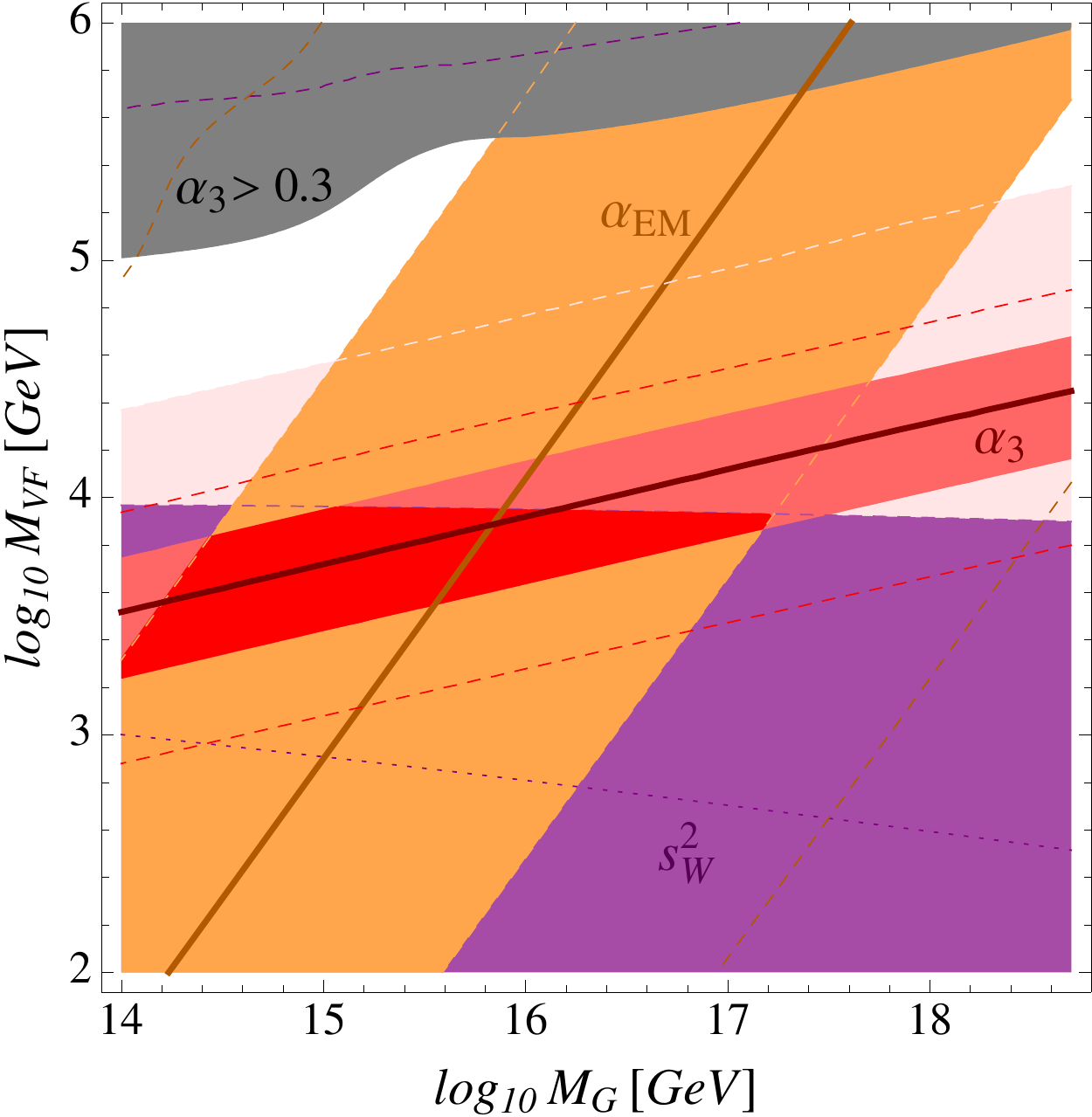}
\caption{Left: contours of constant values of predicted gauge couplings at $M_Z$, $\alpha_1$ (green),  $\alpha_2$ (blue), and $\alpha_3$ (red), as functions of the GUT scale, $M_G$, and the universal mass of fermions from 3 vector-like families, $M_{VF}$, for fixed $\alpha_G = 0.3$. Solid lines represent the central experimental values of three gauge couplings, the shaded regions represent $\pm 10\%$ ranges, and the dashed lines in unshaded areas represent $\pm20\%$ ranges. The lightly shaded area corresponds to a $\pm50\%$ range of $\alpha_3$. In the overlapping (bright red) region,  all three gauge couplings are simultaneously predicted within $10\%$ from the measured values, and the small black area in the red region represents the best fit with all three couplings  within $6\%$ from the measured values. The gray region corresponds to $\alpha_3(M_Z) > 0.3$; $\alpha_3(M_Z) $ becomes non-perturbative very fast with increasing $M_{VF}$ from the value that corresponds to the boundary of this region. Right: the same as in the plot on the left but for $\alpha_{EM}$ (orange),  $s^2_W$ (purple), and $\alpha_3$ (red). The dotted purple line represents  $s^2_W$ being $-5\%$ from the central value.}
\label{fig:M_G_vs_M_VF_3VF}
\end{figure}

The low sensitivity of predicted values of gauge couplings to fundamental parameters is demonstrated in Fig.~\ref{fig:M_G_vs_M_VF_3VF} (left). It shows a large region of the GUT scale, $M_G$, and the universal mass of fermions from 3 vector-like families, $M_{VF}$,  
from which the values of gauge couplings at the EW scale are simultaneously predicted within $10\%$ from the measured values. It also shows the best fit that predicts all couplings within $6\%$. The GUT scale is the best motivated between $10^{15}$~GeV and $10^{17}$~GeV with the best fit close to $10^{16}$~GeV.  For completeness,  a similar plot for $\alpha_{EM}$,   $s^2_W$, and $\alpha_3$ is presented in Fig.~\ref{fig:M_G_vs_M_VF_3VF} (right). However, as we will see from the discussion of threshold corrections in the next subsection, the plot on the left for $\alpha_1$,  $\alpha_2$, and $\alpha_3$ is more indicative of the best motivated values of $M_G$ and $M_{VF}$.

In order to understand the sensitivity of the EW scale values of gauge couplings to fundamental parameters quantitatively, it is instructive to estimate separate contributions to $\alpha_{1,2,3} (M_Z)$  from $\alpha_G$, $M_G$, and $M_{VF}$. Note that the values of $\alpha_{1,2,3}^{-1} (M_Z)$ are approximately  59, 30, and 8.4, respectively. 
From Eqs.~(\ref{eq:sol_1loop}) and (\ref{eq:prediction_1+2loop}), we see that $\alpha_G \gtrsim 0.3 $ contributes less than $\sim10\%$  to the EW scale values of gauge couplings. It is the least important parameter.
Plots in Fig.~\ref{fig:M_G_vs_M_VF_3VF}  for any $\alpha_G > 0.3$ would look almost identical. Increasing $\alpha_G$ moves all the contours slightly to the right. 
The largest contribution to EW scale values of couplings originates from  the $1/(2\pi) \ln (M_G/M_Z) \simeq 5.2$ term multiplied by corresponding beta function coefficients.

The second largest contribution to gauge couplings comes from masses of vector-like fermions. The IR fixed point predictions for the gauge couplings at the EW scale, obtained from  Eq.~(\ref{eq:sol_1loop}) for $\alpha_{1,2}$  with $\alpha_{1,2} (M_G) = \alpha_G$, and from  Eq.~(\ref{eq:sol_2loop}) for $\alpha_{3}$, are modified by threshold corrections $T_i$:
 \begin{equation}
 \alpha_i (M_Z) \; \to \;  \frac{\alpha_i (M_Z)} {1 - \alpha_i (M_Z) T_i}  ,
 \label{eq:a+T}
  \end{equation}
 that depend on masses of the extra vector-like fermions.  These threshold effects  are well approximated
by the leading logarithmic corrections:\footnote{These corrections correspond to removing one loop contributions of vector-like fermions from Eqs.~(\ref{eq:sol_1loop}) and  Eq.~(\ref{eq:sol_2loop}) below their mass. It is an excellent approximation for  $\alpha_{1,2}$ and sufficient  approximation for $\alpha_{3}$ since, for the IR value of $\alpha_{3}$,  the 1-loop term in the RG equation dominates.}
   \begin{eqnarray}
T_i  &= &  \frac{1}{2 \pi}  \sum_{f}  b_i^f  \ln \frac{M_f}{M_Z} ,
\label{eq:Ti}
  \end{eqnarray}
where $b_i^f$ is the contribution  of a given fermion $f$, with mass $M_f$, to the corresponding beta function coefficient~\cite{Machacek:1983tz}.  For particles originating from vector-like families, these contributions, summarized in Table~\ref{tab:b}, are identical to contributions from fermions in the standard model. The contribution from the complete family is identical to all three beta function coefficients and equal to 4/3 for a chiral family, and 8/3 for a vector-like pair ($16 + \overline{16}$ in the SO(10) language).

\begin{table}[t]
\caption{Quantum numbers and contributions to beta function coefficients of particles from extra vector-like families.  The names are chosen to mimic those of the standard model particles with the same quantum numbers. For each particle, there is a corresponding vector-like partner, and its contributions to the beta function coefficients are identical. The $b_1$ coefficients correspond to the SU(5) normalization of the hypercharge. }
\begin{center}
\begin{tabular}{|c|ccccc|ccc|}
\hline
\hspace{0.2cm}  Particle \hspace{0.2cm}  & \hspace{0.2cm}  $SU(3)$  \hspace{0.2cm}&  $\times$  &  \hspace{0.2cm} $SU(2)$  \hspace{0.2cm} & $\times$ & \hspace{0.2cm} $U(1)$ \hspace{0.2cm}  & $ \hspace{0.4cm} b_3 \hspace{0.4cm} $ & $\hspace{0.2cm} b_2\hspace{0.2cm}$ & $\hspace{0.4cm}b_1 \hspace{0.4cm}$ \\
\hline 
Q   &  3 &&  2 && 1/6 & 2/3 &  1 &  1/15\\
U   &  $\bar 3$ && 1  &&  -2/3 & 1/3 &  0 &  8/15\\
E   &  1 && 1 && 1 & 0 &  0 &  2/5\\
L   &  1  && 2  && -1/2 & 0 &  1/3 &  1/5\\
D   &  $\bar 3$ && 1  && 1/3 & 1/3 &  0 &  2/15\\
\hline
\end{tabular}
\end{center}
\label{tab:b}
\end{table}%

The  correction to $\alpha_3$ of about $+40\%$   is crucial in order  to reproduce the measured value. As can be seen  in Figs.~\ref{fig:thresholds} and ~\ref{fig:M_G_vs_M_VF_3VF}, it is indeed $\alpha_3$ that determines $M_{VF} \simeq 10^4$ GeV and, consequently, $M_G \simeq 10^{16}$~GeV. The other two couplings are within 10\% from measured values in much larger ranges of $M_{VF}$ and  would actually prefer smaller $M_{VF}$ and $M_G$. 

Out of the three parameters, the EW scale values of gauge couplings are the most sensitive to changes in $M_{VF}$. However, since $M_{VF}$ is only responsible for at most $\sim 40\%$ of the EW scale values of couplings, the overall sensitivity is still very small. Most of the EW scale values of couplings originate from the IR fixed point. 
Since no precise cancellations between separate contributions  are required, there are  large ranges of fundamental  parameters from which the predicted values of gauge couplings at the EW scale are close to observed values.

\begin{figure}[t]
\includegraphics[width=2.9in]{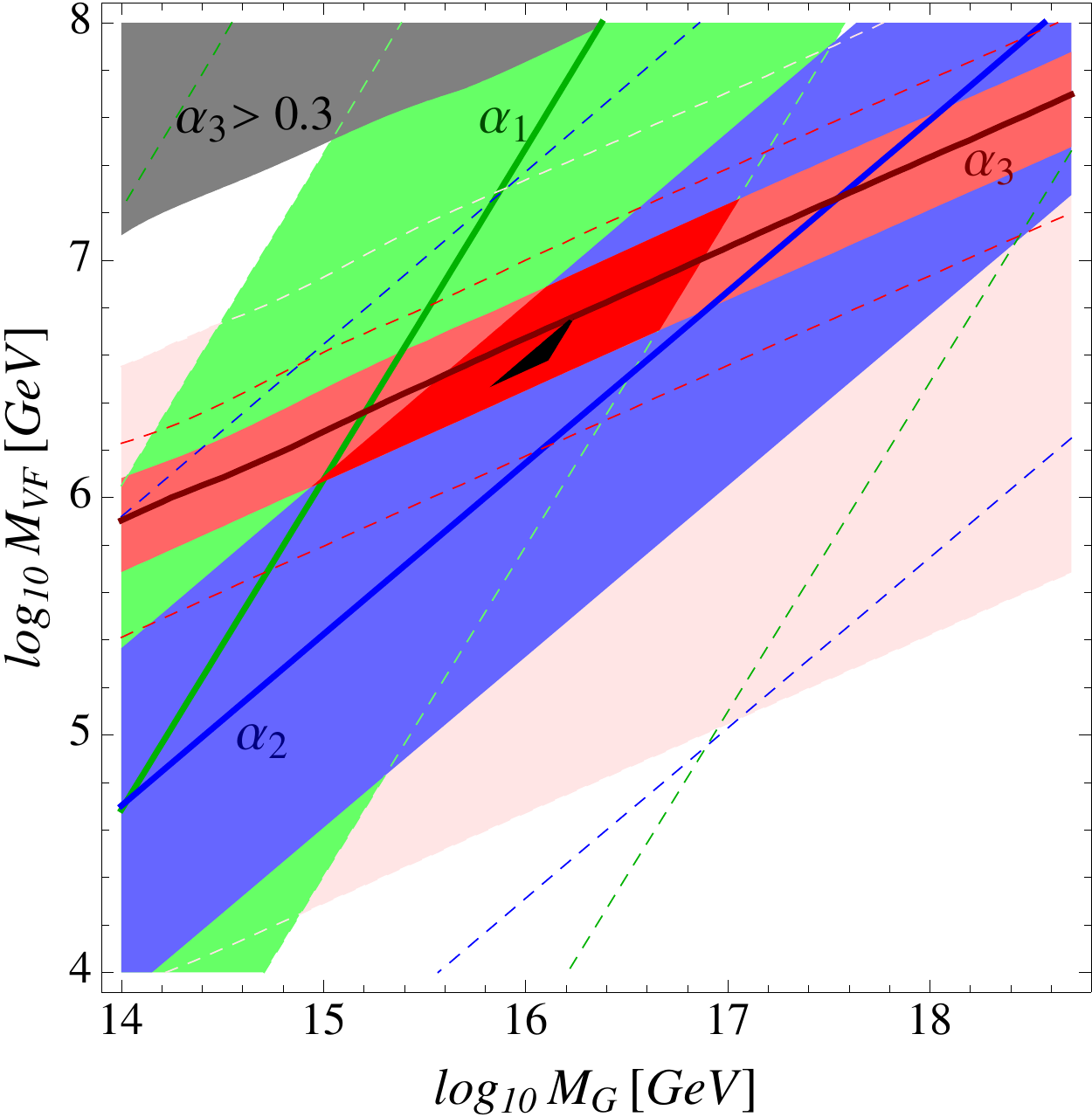} \hspace{0.5cm}
\includegraphics[width=2.9in]{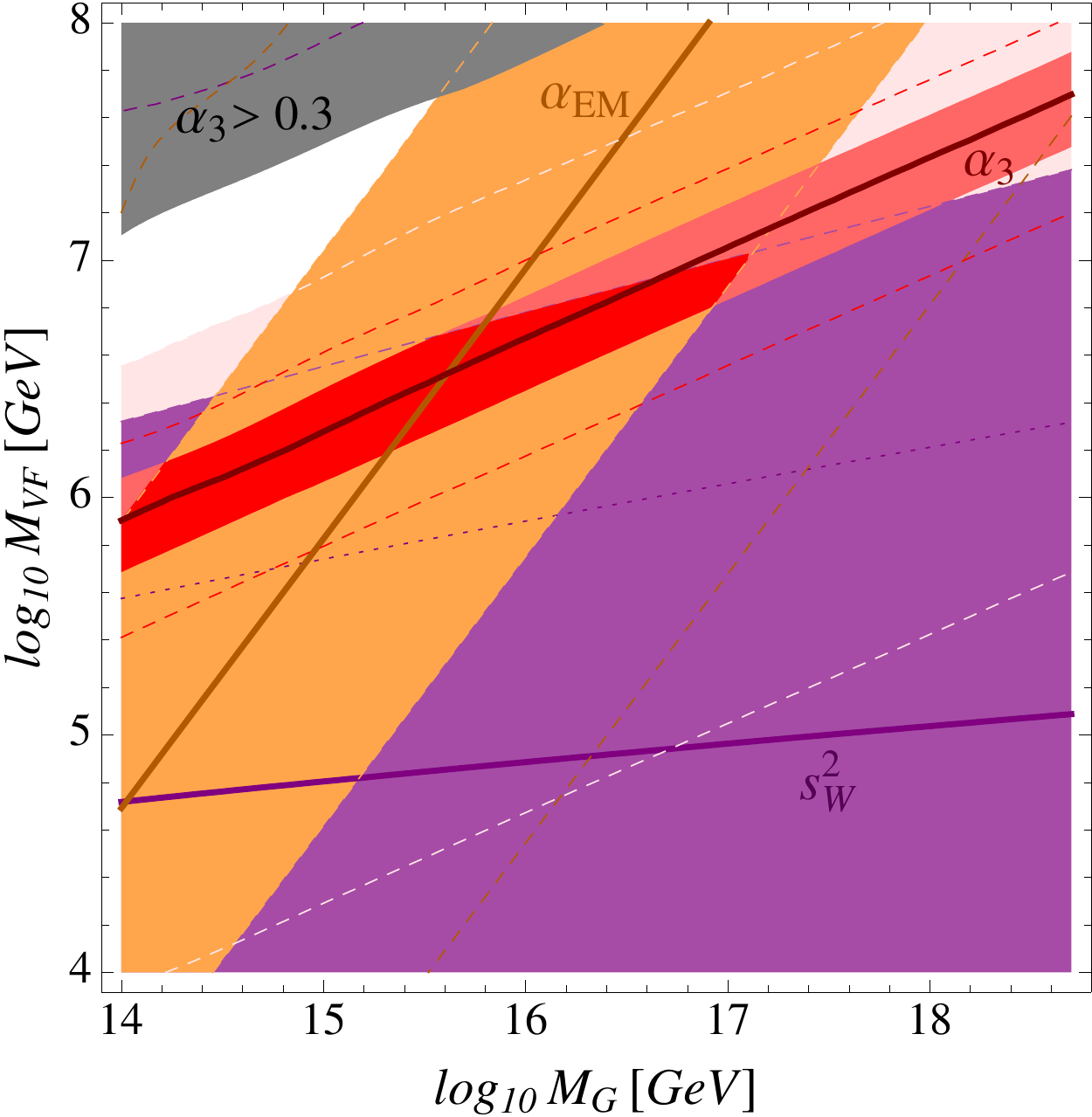}
\caption{The same as in Fig.~\ref{fig:M_G_vs_M_VF_3VF} but for the SM extended by 4 vector-like families. }
\label{fig:M_G_vs_M_VF_4VF}
\end{figure}

The standard model extended by 4 vector-like families (SM+4VFs) allows for insensitive unification of gauge couplings in a similar way as the SM+3VF. Predicted values of gauge couplings at $M_Z$ as functions of the GUT scale and the universal mass of fermions from 4 vector-like families for fixed $\alpha_G = 0.3$ are shown in Fig.~\ref{fig:M_G_vs_M_VF_4VF}. There are, however, notable differences from the SM+3VFs case. First of all, the common mass of vector-like families moves to $\sim 10^6 - 10^7$ GeV. This is easily understood from the fact that more matter makes gauge couplings run faster, and thus the VFs must stop contributing to RG evolution at a higher scale; otherwise, the EW scale values of gauge couplings would be too small. Second of all, the 1-loop IR fixed point value of $\sin^2 \theta_W $ is 0.234 which is larger than in the SM+3VFs case and actually very close to the measured value. Overall, this however does not make the predictions much better than in the SM+3VFs case, since $\alpha_3$ requires $M_{VF}$ larger than the one needed to reach the measured value of $\sin^2 \theta_W $. Finally, as a result of larger  masses of VFs  required in the SM+4VFs scenario, the sensitivity of EW scale values of gauge couplings to fundamental parameters  increased, which is visible in  Fig.~\ref{fig:M_G_vs_M_VF_4VF} as narrower $10\%$ bands compared to those in Fig.~\ref{fig:M_G_vs_M_VF_3VF} corresponding to the case of SM+3VFs.

It is easy to extrapolate to a larger number of VFs. Increasing the number of VFs requires larger $M_{VF}$ closer and closer to the GUT scale. The sensitivity of predicted values of gauge couplings to fundamental parameters is increasing and approaching the sensitivity in the SM.

In the SM extended by 1 or 2 vector-like families the predictive power is  lost, since the unified gauge coupling is small and its specific value is crucial for predictions for gauge couplings at the EW scale in a similar way as in the SM. The  difference form the SM is that 
 the exact unification of gauge couplings is 
now possible with split masses of VFs. We will include these solutions as a curiosity in the next subsection.

\subsection{Threshold effects of vector-like fermions}

Let us now turn our attention to precise predictions for gauge couplings rather than a $\sim$10\% agreement. For this we need to consider threshold effects from splitting  masses of VFs.

The necessity to split masses of particles from extra 3VFs is indicated in Fig.~\ref{fig:thresholds} (left) by slightly different  scales at which the RG evolutions of gauge couplings in the SM and SM+3VFs cross. For the example in this figure the  crossing scales  for $\alpha_1$, $\alpha_2$, and $\alpha_3$ are $M_1\simeq 100$ TeV, $M_2 \simeq 1$ TeV, and $M_3 \simeq 10$ TeV. 
These scales determine threshold corrections $T_i = (4/\pi) \ln (M_{i}/M_Z)$, see Eq.~(\ref{eq:Ti}), required for gauge coupling unification. Any spectrum that leads to required threshold corrections will reproduce the measured values of gauge couplings.

The crossing scales are increasing with increasing  $M_G$ and depend very little on $\alpha_G$ for $\alpha_G \gtrsim 0.3$. For different values of $M_G$, they can be read out of Fig.~\ref{fig:M_G_vs_M_VF_3VF} (left) as corresponding values of $M_{VF}$ for which we obtain the measured value of given gauge coupling. Similarly, in the case of SM+4VFs, the values of $M_{1,2,3}$ can be read out of Fig.~\ref{fig:M_G_vs_M_VF_4VF} (left). For values of $M_G$ not shown, or for other scenarios, the crossing scales can be easily calculated from RG equations as functions of $M_G$ and $\alpha_G$. 

In general, for $N$ pairs of vector-like families, once we know values of crossing scales $M_{1,2,3}$ for chosen GUT scale, the masses of fermions must satisfy:
 \begin{eqnarray}
\frac{4N}{3\pi}  \ln \frac{M_{3}}{M_Z}  &=&   \frac{1}{ \pi}  \sum_{i = 1}^{N}  \left( b_3^Q  \ln \frac{M_{Q_i}}{M_Z} 
+ b_3^U  \ln \frac{M_{U_i}}{M_Z} + b_3^D  \ln \frac{M_{D_i}}{M_Z} 
\right),  \\
\frac{4N}{3\pi}  \ln \frac{M_{2}}{M_Z}   &=&   \frac{1}{ \pi}  \sum_{i = 1}^{N}  \left( b_2^Q  \ln \frac{M_{Q_i}}{M_Z} 
 + b_2^L  \ln \frac{M_{L_i}}{M_Z}
\right), \\
\frac{4N}{3\pi}  \ln \frac{M_{1}}{M_Z}   &=&  \frac{1}{ \pi}  \sum_{i = 1}^{N}  \left( b_1^Q  \ln \frac{M_{Q_i}}{M_Z} 
+ b_1^U  \ln \frac{M_{U_i}}{M_Z} + b_1^D  \ln \frac{M_{D_i}}{M_Z} + b_1^L  \ln \frac{M_{L_i}}{M_Z} + b_1^E  \ln \frac{M_{E_i}}{M_Z}
\right) ,
  \end{eqnarray}
  in order to get exact gauge coupling unification at given GUT scale.
In the case of universal masses of particles with the same  quantum numbers, {\it e.g. } $M_{Q_1}=M_{Q_2}= \dots = M_{Q_N} \equiv M_{Q}$, these mass rules can be written in a simple form:
 \begin{eqnarray}
M_{i}^{4/3} &=& \prod_{F = Q,U,D,L,E} M_{F}^{b_i^F} , \quad \quad  i = 1,2,3. \label{eq:Mrule} 
  \end{eqnarray}
Inserting the beta function coefficients from Table~\ref{tab:b}, we find:
  \begin{eqnarray}
M_{3}^{4} &=& M_{Q}^{2}  M_{U} M_{D} ,\label{eq:M3rule} \\
M_{2}^{4} &=& M_{Q}^{3}  M_{L},\\
M_{1}^{20} &=& M_{Q}  M_{U}^{8} M_{D}^{2}  M_{L}^{3}  M_{E}^{6}. \label{eq:M1rule} 
  \end{eqnarray}  
  These formulas hold for any number of complete vector-like families, only the values of $M_{1,2,3}$ depend on the specific scenario.
In the case of non-universal masses of particles with the same quantum numbers the above formulas are still valid with the replacement:
 \begin{equation}
 M_{F} \equiv  \left( M_{F_1} M_{F_2} \dots M_{F_N}\right)^{1/N}, \quad \quad F = Q,U,D,L,E.
 \label{eq:split}
  \end{equation}

From Eqs.~(\ref{eq:M3rule}) - (\ref{eq:split}) we can immediately see that splitting fermions with the same quantum numbers does not help to find a solution if a solution does not exist with universal masses. Thus, it is sufficient to assume universal masses, $M_{F}$, of particles with the same quantum numbers, and Eqs.~(\ref{eq:M3rule}) - (\ref{eq:M1rule}) classify possible solutions. In addition, for each solution with universal masses, there are other solutions with split masses, and the only constraint is that their geometric mean is the universal mass needed for a given solution.

There are many solutions available, since we have 5 different masses that have to satisfy three conditions (\ref{eq:M3rule}) - (\ref{eq:M1rule}) . However, it is not guaranteed that for a given GUT scale there is a  phenomenologically viable solution. Clearly, the crossing scales have to be above the EW scale, and even then the solution might require new fermions below experimental limits or some fermions above the GUT scale or the Planck scale. 

\begin{figure}[t]
\includegraphics[width=4.in]{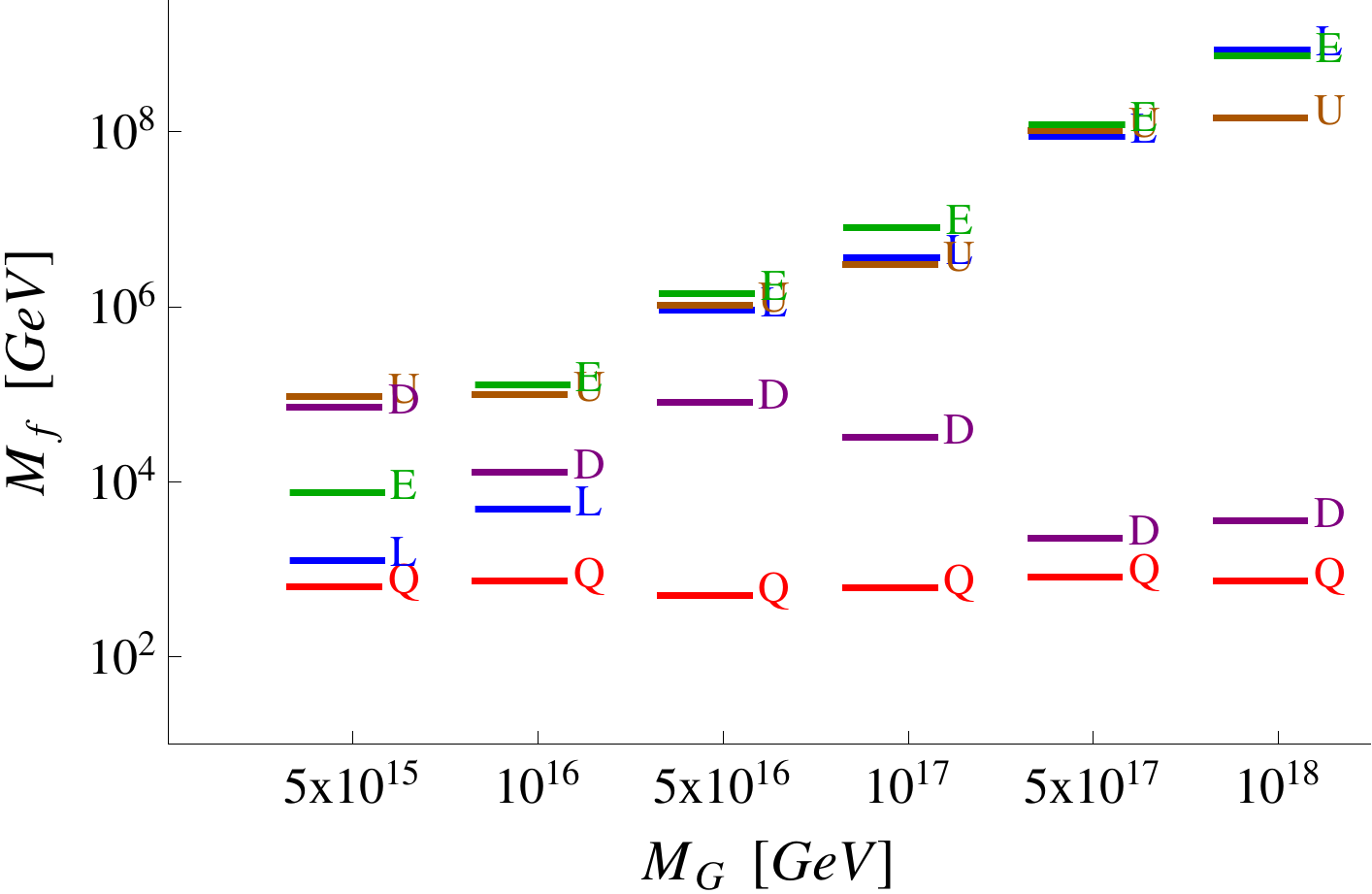}
\caption{Masses of vector-like fermions leading to exact gauge coupling unification as functions of the GUT scale in the case of SM+3VFs. The universal mass for particles with the same quantum numbers is assumed. The value of  $\alpha_G$ is fixed to 0.3.
  Smaller values of  $M_G$ (not shown) can still be consistent with gauge coupling unification for smaller $\alpha_G$. The spectrum shown is just an example, it is not unique.}
\label{fig:Spectrum3VF}
\end{figure}

Representative examples of the spectrum for various values of $M_G$ in the case of SM+3VFs are given in Fig.~\ref{fig:Spectrum3VF}. The value of  $\alpha_G$ is fixed to 0.3; however, the spectrum is not very sensitive to this choice as previously discussed. The spectrum shown is just an example, motivated by the smallest splitting between masses required, and it is not unique. A specific example with exact numerical values was also given in Ref.~\cite{Dermisek:2012as}.  The GUT scale motivated by the lowest splitting required between masses of vector-like fermions is at $\sim 10^{16}$ GeV in agreement with what is suggested in Fig.~\ref{fig:M_G_vs_M_VF_3VF} (left), and the masses are split between $\sim$1 TeV and $\sim$100 TeV. 

There is a lower bound on the possible GUT scale at $\sim 10^{15}$ GeV. For smaller $M_G$ the crossing scale $M_2$ is too small, see Fig.~\ref{fig:M_G_vs_M_VF_3VF} (left), and thus a phenomenologically viable solution does not exist. With increasing $M_G$, the splitting of fermion masses is increasing, which can also be inferred from a larger splitting of crossing scales. The GUT scale can be as high as the Planck scale. However, in that case the masses of vector-like fermions have to be split over 6 orders of magnitude.

Note that quark doublets, $Q$, are typically predicted at $\sim$1 TeV. The preference for $Q$ being  the lightest of vector-like fermions can be understood from $M_2 < M_{1,3}$. However, there are also solutions with $L$ being the lightest. Keep in mind, however, that these masses represent  geometric means of masses of particles with the same quantum numbers. Therefore, when considering split masses of fermions with the same quantum numbers, any fermion can  be the  lightest one and as light as current experimental limits.

\begin{figure}[t]
\includegraphics[width=4.in]{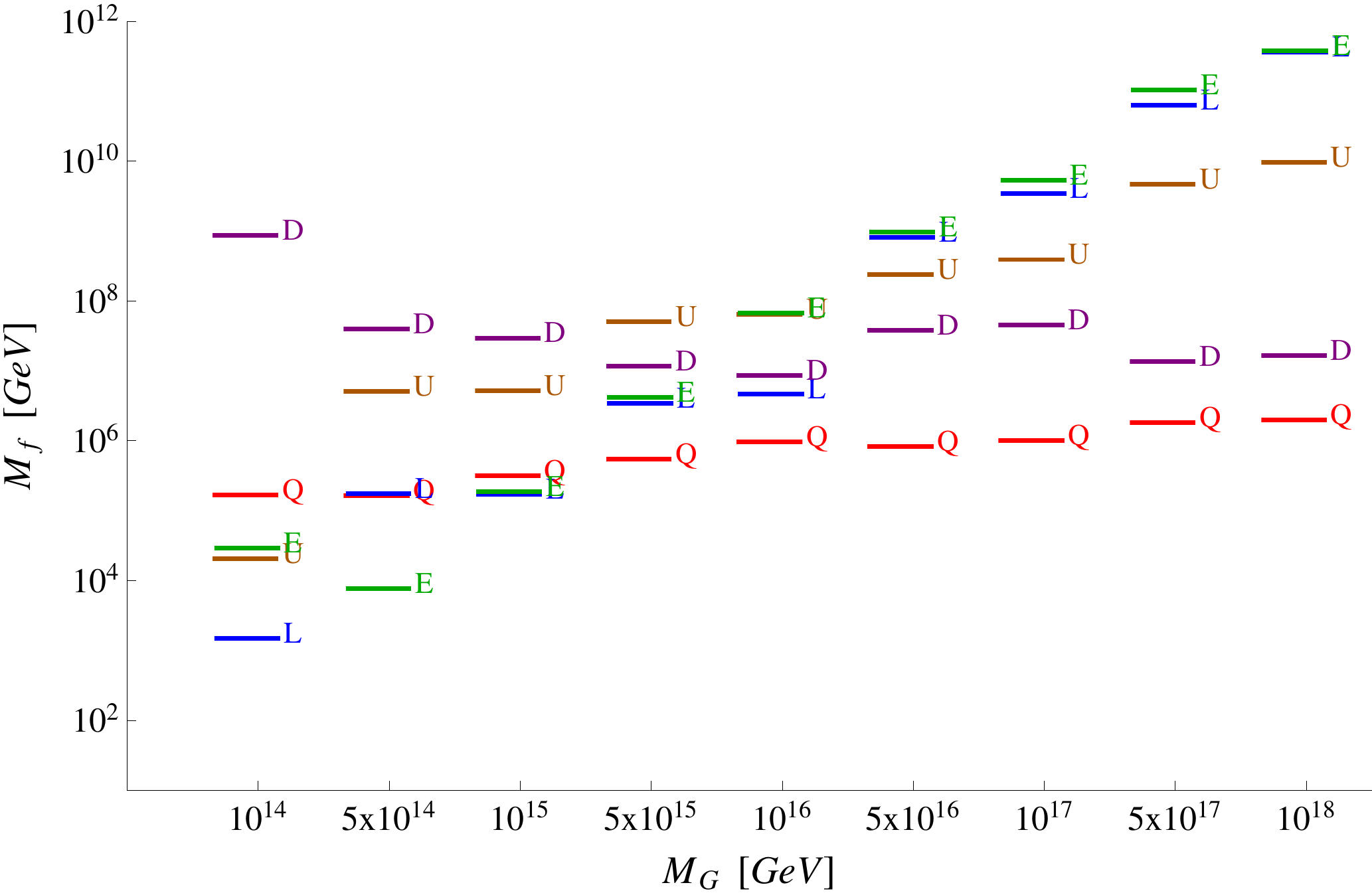}
\caption{The same as in Fig.~\ref{fig:Spectrum3VF} but in the case of SM+4VFs. }
\label{fig:Spectrum4VF}
\end{figure}

For SM+4VFs, examples of the spectrum are given in Fig.~\ref{fig:Spectrum4VF}. The main features are very similar to the case of  SM+3VFs. The GUT scale motivated by the lowest splitting required between masses of vector-like fermions is also at $\sim 10^{16}$ GeV, in agreement with what is suggested in Fig.~\ref{fig:M_G_vs_M_VF_4VF} (left), and about two orders of magnitude splitting of masses of vector-like fermions is required. The main difference from the SM+3VFs case is that the spectrum shifted to $10^6 - 10^8$ GeV.

\begin{figure}[t]
\includegraphics[width=3.5in]{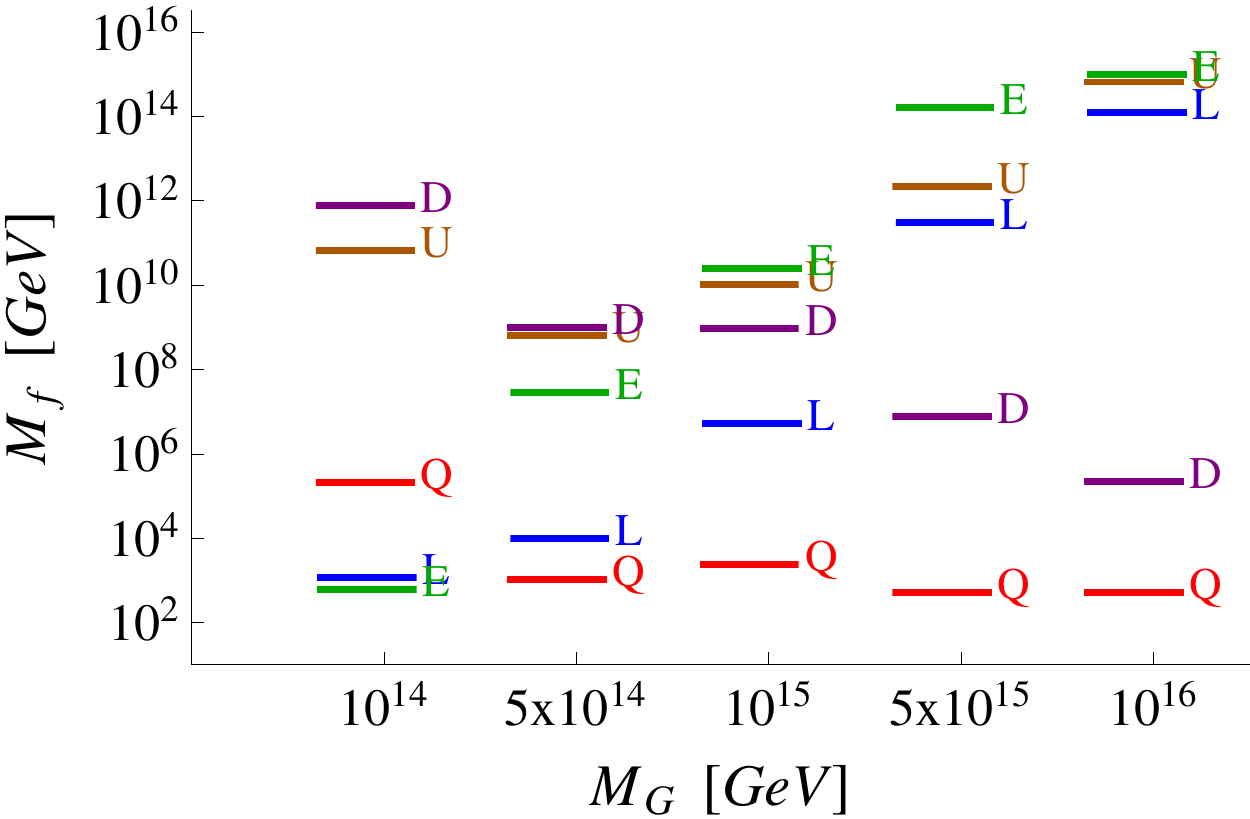}
\caption{The same as in Fig.~\ref{fig:Spectrum3VF} but in the case of SM+1VF. In this case, values of  $\alpha_G$ are optimized for given GUT scale, and are 
  close to  0.03 for all $M_G$ shown. For smaller or larger values of  $M_G$, the unification is not possible for any spectrum.}
\label{fig:Spectrum1VF}
\end{figure}

\begin{figure}[t]
\includegraphics[width=5.in]{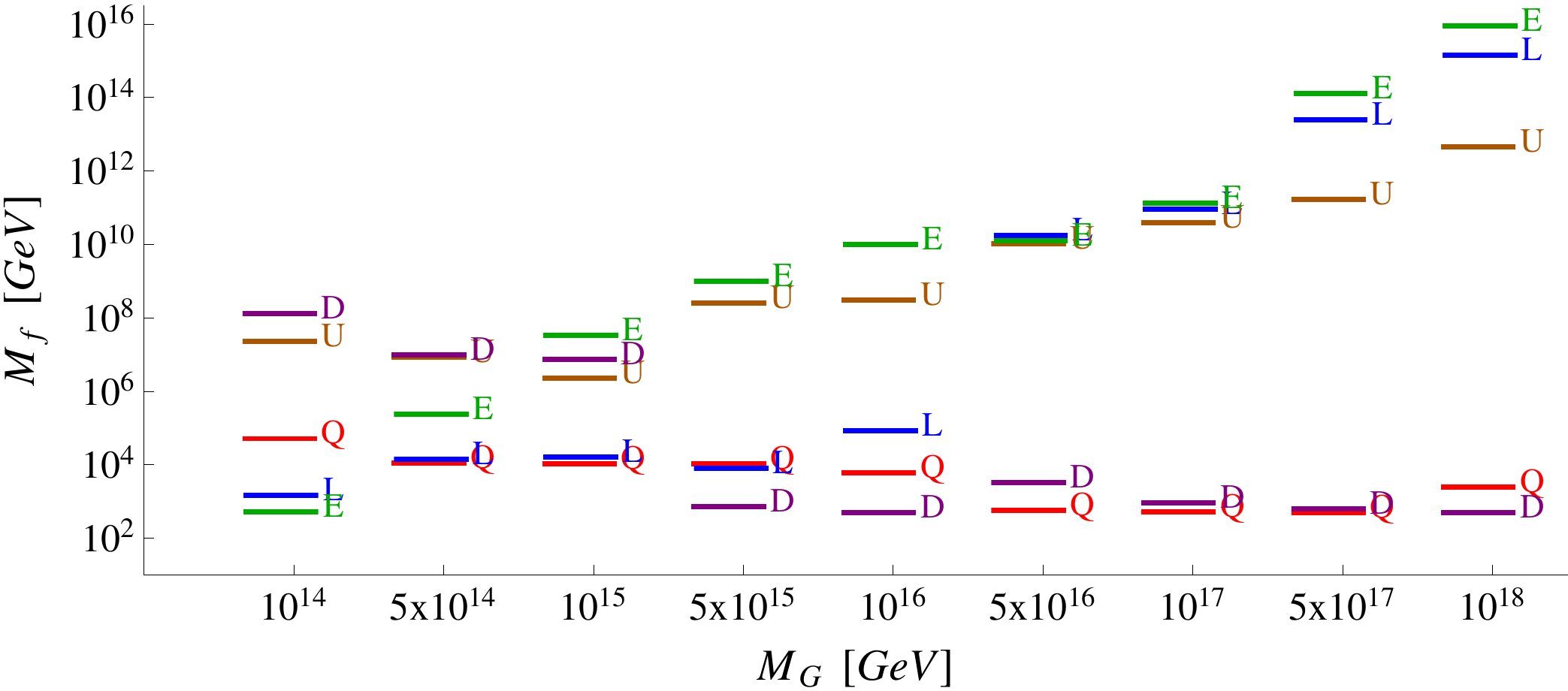}
\caption{The same as in Fig.~\ref{fig:Spectrum3VF} but in the case of SM+2VF. In this case values of  $\alpha_G$ are optimized for given GUT scale, and 
 vary between  0.042 and 0.048. Smaller values of  $M_G$ (not shown) can still be consistent with gauge coupling unification.}
\label{fig:Spectrum2VF}
\end{figure}

For completeness, we also include examples of the spectrum needed for exact gauge coupling unification in the case of SM+1VF in Fig.~\ref{fig:Spectrum1VF} and SM+2VFs in Fig.~\ref{fig:Spectrum2VF}. For these cases, the EW scale values  of gauge couplings are highly sensitive to $\alpha_G$. Thus  $\alpha_G$ in these examples is not fixed but rather optimized for the given $M_G$. In both cases, the exact unification can be achieved even in the region consistent with limits on proton lifetime. However, the required splitting between masses of vector-like fermions is sizable. 

\subsection{Generalization of mass rules to other extensions of the SM}

The mass rules we have just derived can be generalized to any extension of the SM.
  The existence of crossing scales is a necessary  condition for achieving gauge coupling unification in a given model. This follows from the fact that  
 integrating out extra fields above the EW scale can only increase gauge couplings at the EW scale. Therefore, 
  values of predicted couplings at the EW scale without considering the mass effect of extra matter fields have to be smaller than the measured values. 
  The crossing scales depend only on $\alpha_G$ and $M_G$.
  Thus, requiring that the crossing scales exist  leads to  limits on possible values of the  GUT scale and  $\alpha_G$. 
 
  For chosen  $\alpha_G$ and $M_G$, we can  find the crossing scales $M_{1,2,3}$ for all three gauge couplings. If one loop RGEs are good approximations, these crossing scales can be easily found by applying Eq.~(\ref{eq:sol_1loop}) separately between $M_Z$ the $M_{i}$ scales using the SM beta function coefficients, $b_i^{SM}$, starting with the observed values of $\alpha_{i, exp}(M_Z)$, and between  $M_i$ and $M_G$ scales using beta function coefficients in the given extension, $b_i$, assuming  gauge couplings exactly unify. We get:
     \begin{eqnarray}
\ln \frac{M_i}{M_Z} &=& \frac{2\pi}{b_i - b_i^{SM}}  \left( -\alpha_{i, exp}^{-1}(M_Z) +  \alpha_G^{-1} + \frac{b_i }{2\pi} \ln \frac{M_G}{M_Z} \right) .
\label{eq:crossing}
  \end{eqnarray}
 The meaning of crossing scales is the same as in extensions of the SM with VFs; namely, they represent the threshold corrections, $T_i = (b_i - b_i^{SM})/(2\pi) \ln (M_{i}/M_Z)$, that masses of extra particles must generate in order to
reproduce the measured values of gauge couplings starting from the given $\alpha_G$ and $M_G$.
The rest follows what we did for complete VFs. 
 Once we know values of crossing scales $M_{1,2,3}$,  in order to get exact gauge coupling unification, the masses of extra particles must satisfy: 
   \begin{eqnarray}
M_{i}^{(b_i -b_i^{SM})} &=& \prod_{F} M_{F}^{b_i^F}  , \quad \quad  i = 1,2,3, \label{eq:MruleGen} 
  \end{eqnarray}
where the product is over all extra fermions (or scalars)  charged under a given gauge symmetry. For a vector-like pair of fermions, the corresponding mass on the right-hand side appears twice. 
  As in the case of complete VFs, it is sufficient to consider the universal mass of all particles with the same quantum numbers. The universal mass  that enters Eq.~(\ref{eq:MruleGen}) represents their geometric mean.

For any model with an arbitrary particle content, the crossing scales (\ref{eq:crossing}) as functions of $\alpha_G$ and $M_G$ together with the mass rules (\ref{eq:MruleGen}) classify all the solutions consistent with gauge coupling unification  in terms of physical masses of extra particles.

Let us illustrate the usefulness of  crossing scales and the mass rules on one example. Let us ask if there is any spectrum of extra particles in the SM extended by one vector-like family that leads to exact gauge coupling unification for $M_G = 10^{16}$ GeV.  This choice  corresponds to one of the points in Fig.~\ref{fig:Spectrum1VF}, and so we already have the answer we can compare with.
However, this answer is obtained by a fairly complicated numerical procedure  that iteratively solves coupled differential equations with masses of extra vector-like fermions  varied untill the EW scale values of gauge couplings are precisely reproduced. Using our method, we can get the basic features of the required spectrum fast. 

For  $\alpha_G = 0.0286$ that corresponds to the given example in Fig.~\ref{fig:Spectrum1VF}, the crossing scales $M_{1,2,3}$, easily calculated from Eq.~(\ref{eq:crossing}), are $1 \times 10^{13}$ GeV, $7 \times 10^4$ GeV, and $2 \times 10^6$ GeV. This immediately tells us that there will be more than 8 orders of magnitude splitting between masses required. Knowing the crossing scales, we can easily see the basic features of the spectrum that will work. From Eqs.~(\ref{eq:MruleGen}), which in this case are the same as Eqs.~(\ref{eq:M3rule}) - (\ref{eq:M1rule}), we see that $M_Q$, which heavily weighs on $M_2$, should be less than $M_2$, while everything with large hypercharge (especially $E$ and $U$)  should be above $M_1$ in order to find a solution. For a specific example, one can choose two masses and calculate the rest of the spectrum from  Eqs.~(\ref{eq:M3rule}) - (\ref{eq:M1rule}).  Given a large splitting between $M_2$ and $M_1$ in this case, it would be easiest to choose the masses of Q and  U as a starting point. Once we have one solution, varying the starting masses of Q and  U and calculating the rest of the masses from Eqs.~(\ref{eq:M3rule}) - (\ref{eq:M1rule})  will give us all possible solutions for the given $\alpha_G$ and $M_G$. This procedure can be repeated for any $\alpha_G$ and $M_G$, or the solutions can be plotted as functions of these variables.

\section{Discussion}
\label{sec:discussion}

So far, we have only considered constraints on the GUT scale and masses of vector-like fermions from gauge coupling unification. In order for this scenario to be easily embedded into simple grand unified theories, based on $SU(5)$ or $SO(10)$, the constraints on proton lifetime and the stability of the EW minimum of the Higgs potential should be satisfied.

  The most stringent limits on proton lifetime  come from Super-Kamiokande. For the dominant decay mode from  dimension-6  operators, the limit is 
  $\tau (p \to \pi^0 e^+)  > 1.4 \times 10^{34}$ yrs~\cite{Hewett:2012ns}.  Assuming naively, that the proton lifetime is $\tau_p \sim M^4_G/(\alpha_G^2 m_p^5)$, where $m_p$ is the mass of the proton, this limit translates into the lower bound on the GUT scale: $M_G > 1.5 \times 10^{16}$ GeV for $\alpha_G = 0.3$ which we use in our examples. However, the prediction for the proton lifetime is somewhat model dependent (see for example  Refs.~\cite{PDG_GUTs, Hewett:2012ns, Nath:2006ut} and references therein), and so we do not impose the strict limit in the plots we present. In addition, the plots would look very similar for any large value of $\alpha_G$, but the limits would differ. The interested reader can easily impose the limit on any scenario by simple rescaling of the mentioned limit using the formula for the proton lifetime. 
  
  It is interesting to note that the best motivated value of the GUT scale  is in the $\sim$$10^{16}$~GeV range which is basically at the current  limit. It is, however, not possible to make precise predictions without knowing the masses of vector-like fermions. For example, a scenario with  the GUT scale larger by a factor of 3 results in $\sim$2 orders of magnitude enhancement of the proton lifetime but would only require modest changes in the spectrum of vector-like fermions in order to have exact gauge coupling unification. This inability to make precise predictions of GUT scale parameters is a direct consequence of the insensitivity of the EW scale couplings to  GUT scale boundary conditions.   
  
 The RG evolution   of the top Yukawa and   Higgs quartic couplings in the SM+3VFs for  $M_G = 2\times 10^{16}$ GeV and $\alpha_G= 0.3$ was given in Ref.~\cite{Dermisek:2012as}.  The Higgs quartic coupling remains positive all the way to the GUT scale, and thus the electroweak minimum of the Higgs potential is stable.  This result holds in a large range of $M_G$ and $\alpha_G$, especially in the best motivated region. Therefore, these scenarios represent some of the simplest possible extensions of the standard model that can be embedded into grand unified theories, with a sufficiently long lived proton, and a stable EW minimum of the Higgs potential.

  We have not investigated the origin of masses of vector-like fermions needed for gauge coupling unification. This would require additional assumptions about the mechanism that generates them and the scale at which boundary conditions are set. The masses of vector-like fermions may be fundamental lagrangian parameters, or they can originate from Yukawa couplings to one or several additional scalars (singlets under SM gauge symmetry, but possibly charged under family symmetries) that acquire  vacuum expectation values at any scale between the GUT scale and the $M_{VF}$ scale. In addition, vector-like fermions can have non-zero Yukawa couplings to the SM Higgs doublet, which add another layer of complexity by contributing to the physical masses and possibly significantly affecting the RG evolution of  other parameters that directly determine their masses.   The study of gauge coupling unification is, to a large extent, unaffected by  these assumptions, only the physical masses of particles  matter in the leading order. Fundamental lagrangian masses would not affect the running of gauge couplings  at all, and the Yukawa couplings to extra scalars  may only contribute to the RG evolution of gauge couplings at 2-loop level. 
  However, in any specific scenario, the mass rules (\ref{eq:M3rule}) - (\ref{eq:M1rule}) can be evolved to the GUT scale (or other relevant scale), and the freedom to choose some of the masses can be used to search for simple boundary conditions that are consistent with gauge coupling unification.

Finally, 
it is intriguing  to consider a connection with the anthropic solution to the hierarchy problem, or the EW scale~\cite{Agrawal:1997gf, Agrawal:1998xa}. Adding VFs to the SM  makes this possibility more appealing, since in the SM, special values of gauge couplings either at the EW scale or some  high scale have to be selected. In scenarios that we discussed, the EW scale values of gauge couplings close to 
the observed values  are not very special, but rather quite a generic outcome from large ranges of fundamental parameters. For example, in the SM+3VFs case, as far as $M_{VF} < 25$ TeV, for any  $\alpha_G \gtrsim 0.2$, and $M_G$ anywhere between $10^{14}$ GeV and the Planck scale, the predicted values of gauge coupling at the EW scale are always within 50\% of the measured values ($\alpha_{1,2}$ typically well within 20\%). This is indicated by the lightly shaded region in Fig.~\ref{fig:M_G_vs_M_VF_3VF}, and a similar region is indicated in  Fig.~\ref{fig:M_G_vs_M_VF_4VF} for SM+4VFs.
 Furthermore, if the EW scale  and $M_{VF}$ have the same origin, it would also explain the proximity of the QCD scale to the EW scale. The beta function of $\alpha_3$ changes the sign at  $M_{VF}$, and below this scale, it starts running fast toward $\Lambda_{QCD}$.

\section{Conclusions}
\label{sec:conclusions}

We have discussed gauge coupling unification in  models with additional  1 to 4 complete vector-like families. In scenarios with 3 or more vector-like families  the values of  gauge couplings at the electroweak scale  are highly insensitive to the grand unification scale,  the unified gauge coupling, and the masses of vector-like fermions. Their observed values can be mostly understood from  infrared fixed point behavior. Starting with a large (but still perturbative) unified gauge coupling at a high scale,  
the values of gauge couplings at lower energies are determined only by the particle content of the theory and how far from the GUT scale we measure them. Since the exact value of $\alpha_G$ becomes irrelevant, instead of one prediction of the conventional unification, we have two predictions for ratios of gauge couplings. These predictions are modified at the $M_{VF} $ scale, where the vector-like fermions are integrated out, and below this scale,  gauge couplings run according to the usual RG equations of the standard model.

 Assuming first a common mass of vector-like fermions, $M_{VF}$,  
 we showed predictions for three gauge couplings at the EW scale as functions of $M_G$ and $M_{VF}$. We found that the observed values of gauge coupling are reproduced with good precision from a large range of parameters.  Especially, $M_G$ can be varied over several orders of magnitude while having all three gauge couplings within 20\% from observed values. The  best fit, which  predicts all three couplings within 6\% from measured values, suggests $M_G \sim 10^{16}$ GeV,  $M_{VF} \simeq 10^4$ GeV in the case of SM+3VFs, and $M_{VF} \simeq 10^6 - 10^7$ GeV in the case of SM+4VFs.

The best motivated GUT scale, $\sim$$10^{16}$ GeV, predicts a proton lifetime close to current limits. 
 However, due to insensitivity of the predicted EW scale values of gauge couplings to GUT scale parameters, no sharp predictions can be made without knowing the spectrum of vector-like fermions. In addition, it was previously shown that the Higgs quartic coupling remains positive all the way to the GUT scale, and thus the electroweak minimum of the Higgs potential is stable.  This result holds in a large range of $M_G$ and $\alpha_G$, especially in the best motivated region.
Therefore, these scenarios represent some of the simplest possible extensions of the standard model that can be embedded into grand unified theories, with a sufficiently long lived proton, and the stable EW minimum of the Higgs potential.

The discrepancies of IR fixed point predictions from observed values can be explained by threshold effects of extra vector-like fermions. We showed examples of the spectrum for the GUT scale varied  between $10^{14}$ GeV and $10^{18}$ GeV. 
We derived simple rules for masses of vector-like fermions required for exact gauge coupling unification. 
In addition, we generalized the mass rules and the method of using crossing scales of evolutions of gauge couplings in the SM and the given extension to classify scenarios consistent with gauge coupling unification to an arbitrary extension of the standard model. 
  The problem of finding all possible mass spectra in a given model consistent with gauge coupling unification is reduced to solving a set of simple algebraic equations that masses of extra particles have to satisfy.

With respect to the sensitivity to  fundamental parameters, the model with 3 extra vector-like families stands out. In the best motivated region, it requires  vector-like fermions with masses of order 1 TeV -- 100 TeV, and thus at least part of the spectrum may be within the reach of  the LHC. 
Notably, quark doublets, $Q$, are typically predicted at $\sim$1 TeV. However, only geometric means of masses of particles with the same quantum numbers are constrained by  gauge coupling unification. Therefore, when considering split masses of fermions with the same quantum numbers, any fermion can  be the  lightest one and as light as current experimental limits. Besides direct production of these particles at the LHC, it may be also possible to observe their effects in a variety of processes. However, they typically affect standard model predictions only through mixing with light fermions, which is highly model dependent. The discussion of gauge coupling unification that we focused on here is negligibly affected by such mixing.


\vspace{0.5cm}
\noindent
{\bf Acknowledgments:} R. D. thanks H.D. Kim and  L. Hall for useful comments and discussions. R. D. also thanks  CERN Theory Institute, the Center for Theoretical Underground Physics and Related Areas (CETUP* 2012), and the Galileo Galilei Institute for Theoretical Physics for hospitality and for partial support during various stages of this work.
This work was supported in part by the  Department of Energy under grant number DE-FG02-91ER40661.



\end{document}